\documentclass[3p,times]{elsarticle}
\usepackage{hyperref}
\usepackage{amsmath}
\usepackage{amssymb}
\usepackage{mathrsfs} 

\usepackage{graphicx}
\usepackage{xcolor}
 
\journal{arXiv}

\begin{document}

\begin{frontmatter}

\title{Fantastic quasi-photon and the symmetries of Maxwell electromagnetic theory,\\ momentum-energy conservation law, and Fermat's principle}
\author{Changbiao Wang}
\ead{changbiao\_ wang@yahoo.com}
\address{ShangGang Group, 70 Huntington Road, Apartment 11, New Haven, CT 06512, USA}

\begin{abstract}
In this paper, I introduce two new concepts (Minkowski quasi-photon and invariance of physical definitions) to elucidate the theory developed in my previous work [Can.\,J.\,Phys.\,93, 1510 (2015)], and to clarify the criticisms by Partanen and coworkers [Phys.\,Rev.\,A 95,\,063850 (2017)].   Minkowski quasi-photon is the carrier of the momentum and energy of light in a medium under the sense of macroscopic averages of light-matter microscopic interactions. I firmly argue that required by the principle of relativity, the definitions of all physical quantities are invariant.  I shed a new light on the significance of the symmetry of physical laws for resolution of the Abraham-Minkowski debate on the momentum of light in a medium.  I illustrate by relativistic analysis why the momentums and energies of the electromagnetic subsystem and the material subsystem form Lorentz four-vectors separately for a closed system of light-matter interactions, and why the momentum and energy of a non-radiation field are owned by the material subsystem, and they are not measurable experimentally.  Finally, I also provide an elegant proof for the invariance of physical definitions, and a clear definition of the Lorentz covariance for general physical quantities and tensors.
\end{abstract}


\end{frontmatter}
\section{Introduction}
\label{s1}
Einstein's 1905 light-quantum hypothesis and energy-mass equivalence equation, combined with Newton's law (momentum = mass multiplied by velocity), suggested a clear definition of the momentum of a photon in \emph{vacuum}, and this definition was strongly supported by the 1923 experiment of photon-electron Compton scattering, and inspired the birth of de-Broglie 1924 momentum-wavelength hypothesis.  However, the momentum of light in a \emph{material medium} has been a perplexing problem in physics for as long as a century.  Various theories have been proposed to resolve this problem; for example, in a recent interesting paper \cite{r1}, Partanen and coworkers argue that the propagation of a light pulse in a medium has to be described by a coupled state of the field and matter, and this coupled state supports a mass density wave (MDW) when driven by the field-dipole forces.  The authors emphasize that by use of mass-polariton (MP) quasiparticle approach and optoelastic continuum dynamics (OCD) approach, respectively, their study results ``provide extremely strong support for the Abraham force density as the only optical force density that is fully consistent with the covariance principle'', and claim that they have founded ``a covariant theory of light propagation in a medium''. \\
\indent In this paper, I would like to indicate that there are fundamental flaws in Partanen-coworkers theory: (i) the criterion of Lorentz four-vector covariance set up by the authors is not mathematically correct, resulting in failure to identify whether their physical model is compatible with Einstein's special relativity; (ii) the invariant definition of refractive index $n$ is not given for a general inertial frame, and thus there is no information that is provided to support the propagation covariance of light in a medium. Based on the analysis of Partanen-coworkers paper and inspired from their physical model, I would like here to propose a new concept, so-called ``Minkowski quasi-photon'', with some fantastic properties exposed. Meanwhile, I firmly argue that to examine the covariance of light propagation in a medium, (a) the definitions of all physical quantities, including the refractive index, must be the same in all inertial frames, and (b) the momentum and energy of light in a medium must constitute a Lorentz four-vector.  \\
\indent The paper is organized as follows.  In Sec.\,\ref{s2}, it is shown why the criterion of Lorentz four-vector covariance set up by Partanen and coworkers is not mathematically correct, and in Sec.\,\ref{s3}, it is indicated why the propagation covariance of light in a medium is not supported by Partanen-coworkers theory.  In Sec.\,\ref{s4}, the fantastic property of Minkowski quasi-photon is exposed, the covariance of propagation direction and speed of light in a medium is formulated, the invariance of physical definitions is argued, and a proof is given of why Partanen-coworkers physical model is not compatible with Einstein's special relativity.  In Sec.\,\ref{s5}, some conclusions are given, and some remarks on the significance of the symmetry of physical laws are made. It is also illustrated by relativistic analysis why the momentums and energies of the electromagnetic (EM) subsystem and the material subsystem constitute Lorentz four-vectors, respectively, for a closed system of light-matter interactions, and why the momentum and energy of a non-radiation field are owned by the material subsystem, instead of the EM subsystem, and they are not measurable experimentally. Finally, an elegant proof is given for the invariance of physical definitions, and what the Lorentz covariance of a general physical quantity and a tensor really means is explained (in footnote 7).  

\section{Problem of the covariance of energy and momentum}
\label{s2}
 Partanen and coworkers argue that the rest mass of MP quasiparticle is given by
\begin{equation}
m_0=n\sqrt{n^2-1}~\hbar\omega/c^2,
\label{eq1}
\end{equation}
and the total energy and momentum are, respectively, given by
\begin{align}
&E_{\mathrm{MP}}=\gamma m_0c^2=\hbar\omega+(\delta m)c^2,
\label{eq2} 
\\
&p_{_{\mathrm{MP}}}=\gamma m_0v=n\hbar\omega/c,
\label{eq3}
\end{align}  
where $\hbar$ is the reduced Planck constant, $\omega$ is the angular frequency, $c$ is the speed of light in vacuum, $\delta m=(n^2-1)\hbar\omega/c^2$,  $v=c/n$, and  $\gamma=(1-v^2/c^2)^{-1/2}$; confer the authors' Eqs.\,(5)-(7) \cite{r1}.  Note that  $v=c/n$  is the propagation velocity of the rest mass $m_0$  of the MP quasiparticle \cite{r1}.

Partanen and coworkers claim that $E_{\mathrm{MP}}$ and $p_{_{\mathrm{MP}}}$ are Lorentz covariant because they fulfill the covariance condition (criterion), given by 
\begin{equation}
E_{\mathrm{MP}}^2-(p_{_{\mathrm{MP}}}c)^2=(m_0c^2)^2. \vspace{2mm}
\label{eq4}
\end{equation} 
Mathematically, what the above Partanen-coworkers criterion means is: For a ``four-component vector''  $(A^0,\mathbf{A})$, if $(A^0)^2-\mathbf{A}^2$ = Lorentz invariant is valid, then $(A^0,\mathbf{A})$ is a Lorentz four-vector.  Unfortunately, this criterion is not correct.

To understand why  $(A^0)^2-\mathbf{A}^2$ = Lorentz invariant  is not a sufficient condition to make  $(A^0,\mathbf{A})$  a four-vector, a simple counterexample is given below. 

It is well known that to keep Maxwell equations invariant in form in all inertial frames, the electric field $\mathbf{E}$ and magnetic field $\mathbf{B}$ must constitute a field strength tensor, which follows \emph{four-tensor} Lorentz transformation, resulting in $(|\mathbf{B}|c)^2-\mathbf{E}^2$ = Lorentz invariant \cite[p.\,82]{r2}. 

According to Partanen-coworkers criterion, $(|\mathbf{B}|c, \mathbf{E})$  is supposed to be a Lorentz four-vector, because it satisfies $(|\mathbf{B}|c)^2-\mathbf{E}^2$ = Lorentz invariant; however, $(|\mathbf{B}|c, \mathbf{E})$  is \emph{never} a four-vector, because it does not follow \emph{four-vector} Lorentz transformation.

The above counterexample clearly disproved the covariance criterion set up by Partanen and coworkers. Thus the covariance problem of the total energy $E_{\textrm{MP}}$  and total momentum $p_{_{\mathrm{MP}}}$ of MP quasiparticle is not resolved in their theory, which  results in a fact that the authors are not aware that their physical model already violates Einstein's special relativity, as shown in footnote 3 of Sec.\,\ref{s4}. 

\section{Problem of the covariance of light propagation} 
\label{s3}
According to the relativity principle, the physical definition of refractive index $n$ should be invariant in all inertial frames.  However Partanen and coworkers did not provide the invariant definition of $n$, although it is used to analyze relativistic covariance throughout their paper. 

The authors indeed defined the refractive index $n$ for an \emph{isotropic} medium in the medium-rest fame (laboratory frame), given by  $n^2=\epsilon_r\mu_r$, with $\epsilon_r$ and $\mu_r$ the relative permittivity and permeability of the medium.  However the authors did not provide the definition in a general moving frame (or in an \emph{anisotropic} medium).  If we don't know the definition of refractive index in a general inertial frame, how do we know ``MP quasiparticle ... propagates through the medium at speed  $v=c/n$", \emph{which holds in all inertial frames}?  Thus, the covariance problem of the propagation (direction and speed) of MP quasiparticle in Partanen-coworkers theory \cite{r1} is not resolved either. 

\section{Quasi-photon and invariance of physical definitions}
\label{s4}
As shown above, Partanen and coworkers failed to prove the four-vector covariance of energy $E_{\mathrm{MP}}=\hbar\omega+(\delta m)c^2$  and momentum $p_{_{\mathrm{MP}}}=n\hbar\omega/c$  for MP quasiparticle, and they did not provide any evidence to support the propagation covariance either; thus Partanen and coworkers did not resolve the covariance problem for their theory.  Accordingly, there is no basis of Partanen-coworkers criticism that my theory \cite{r3,r4,r5,r6}\footnote{
In Ref.\,\cite{r3}, it is indicated that there are two kinds of mass: energy-associated mass  $m_{_E}$, defined through $E=m_{_E}c^2$ (Einstein's energy-mass equivalence equation) \cite{r7}, and momentum-associated mass $m_{\mathbf{p}}$, defined through  $\mathbf{p}=m_{\mathbf{p}}\mathbf{u}$, where  $E$,  $\mathbf{p}$, and $\mathbf{u}$ are, respectively, the particle energy, momentum, and velocity.  For classical particles and photons in vacuum,  $m_{\mathbf{p}}=m_{_E}$ holds, while for Minkowski photon in a medium,  $m_{\mathbf{p}}=n^2\hbar\omega/c^2$ and $m_{_E}=\hbar\omega/c^2$ are valid, with Newton law $|\mathbf{p}|=m_{\mathbf{p}}|\mathbf{u}|=m_{\mathbf{p}}(c/n)=n\hbar\omega/c$ automatically satisfied.  It is also indicated that there are three kinds of momentum-energy four-vectors: $P^{\mu}P_{\mu}=E^2/c^2-\mathbf{p}^2>0$ for classical massive particles, $P^{\mu}P_{\mu}=0$  for photons in vacuum, and $P^{\mu}P_{\mu}<0$  for photons in a medium. Thus $(P^{\mu}P_{\mu})^{1/2}/c$ is real for massive particles and photons in vacuum; however, it becomes imaginary for Minkowski photon in a dielectric medium.
}$^{,}$\footnote{
In Ref.\,\cite{r5}, I showed that Minkowski photon four-momentum is the direct result of Einstein light-quantized electromagnetic (EM) momentum and energy, namely  $\hbar K^{\mu}=N_p^{-1}(\mathbf{g}_{_M},W_{em}/c)$, which holds in all inertial frames, and where $N_p$  is the photon number density in volume, $\mathbf{g}_{_M}=\mathbf{D}\times\mathbf{B}=N_p(\hbar\mathbf{k}_w)$  is the Minkowski EM momentum density vector, and   $W_{em}=0.5(\mathbf{D}\cdot\mathbf{E}+\mathbf{B}\cdot\mathbf{H})=N_p(\hbar\omega)$ is the EM energy density. It is worthwhile to point out that  $\hbar K^{\mu}$  is constructed based on Einstein light-quantum hypothesis (single photon energy $=\hbar\omega$) and the invariance of phase (resulting in the wave four-vector $K^{\mu}$, as shown in footnote 3), while $N_p^{-1}(\mathbf{g}_{_M},W_{em}/c)$  is constructed based on Einstein light-quantum hypothesis [EM energy density $W_{em}=N_p(\hbar\omega$)] and the Lorentz covariance of field-strength four-tensors to keep Maxwell equations invariant in form in all inertial frames.  The invariance of phase is an \emph{independent mathematical requirement} under the Lorentz transformations of the field-strength tensors.  Thus $\hbar K^{\mu}=N_p^{-1}(\mathbf{g}_{_M},W_{em}/c)$  characterizes the self-consistence in mathematics and physics.  In this publication, I also provided the invariant definitions of refractive index and phase velocity. The refractive index is defined as $n\equiv |\mathbf{k}_w/(\omega/c)|$, which can be equivalently expressed as $n=|\mathbf{g}_{_M}/(W_{em}/c)|$. The phase velocity is defined as $\mathbf{v}_{ph}\equiv\mathbf{\hat{n}}(\omega/|\mathbf{k}_w|)=\mathbf{\hat{n}}(\omega/|\omega|)(c/n)$, which can also be expressed as  $\mathbf{v}_{ph}=\mathbf{\hat{n}}(W_{em}/|\mathbf{g}_{_M}|)$. (Note that observed in the medium-rest frame, $\omega>0$  is assumed, which corresponds to a right-handed plane wave.)   Observed in the photon-rest frame [confer Eq.\,(\ref{eq7}) in the present paper], $\mathbf{E}'=0$, $\mathbf{H}'=0$, $\mathbf{D}'\neq 0$, and    $\mathbf{B}'\neq 0$ hold; thus we have $W'_{em}=0$ and  $\mathbf{g}'_{_M}=\mathbf{D}'\times\mathbf{B}'\neq 0$, leading to $n'\equiv |\mathbf{k}'_w/(\omega'/c)|=|\mathbf{g}'_{_M}/(W'_{em}/c)|=\infty$ and $\mathbf{v}'_{ph}\equiv\mathbf{\hat{n}}'(\omega'/|\mathbf{k}'_w|)=\mathbf{\hat{n}}'(W'_{em}/|\mathbf{g}'_{_M}|)=0$.
} ~``neglects the transferred mass  $\delta m$, thus leading to mathematical problems''. 

In my theory, Minkowski photon momentum-energy four-vector is shown to be the unique correct four-momentum of light in a medium, given by $\hbar K^{\mu}=\hbar(\mathbf{k}_w,\omega/c)$, where Planck constant $\hbar$  is a Lorentz invariant, $K^{\mu}$  is the wave four-vector,\footnote{
\emph{Origin of wave four-vector and incompatibility of MP quasiparticle model with special relativity.} For a plane wave in a \emph{moving} non-dispersive, lossless, non-conducting, isotropic, uniform medium, the phase function can be written in a covariant form, given by $\Psi (\mathbf{x},t)=(\omega t-\mathbf{k}_w\cdot\mathbf{x})$ $=g_{\mu\nu}K^{\mu}X^{\nu}$, where $g_{\mu\nu}=g^{\mu\nu}=\mathrm{diag}(-1,-1,-1,+1)$  is the metric tensor, $X^{\mu}=(\mathbf{x},ct)$ is the time-space four-vector, and $K^{\mu}=(\mathbf{k}_w,\omega/c)$ \cite{r5}. (It is should be noted that for the plane wave in such a case, the wave vector $\mathbf{k}_w$ and the angular frequency $\omega$ are independent of $t$ and $\mathbf{x}$, and they are of real number.)  Required by the Lorentz transformations of field-strength tensors to keep Maxwell equations invariant in form in all inertial frames, the phase function $\Psi (\mathbf{x},t)$ must be a Lorentz invariant; accordingly, $K^{\mu}=(\mathbf{k}_w,\omega/c)$ must be a Lorentz four-vector (called ``wave four-vector''), which was actually first shown by Einstein in his 1905 paper when setting up ``theory of Doppler's principle and of aberration'' for a plane wave in free space \cite{r8}, and which is also well presented in textbooks \cite[p.\,125]{r9}\,\cite[p.\,599]{r10}. Thus the Lorentz covariance of $(\mathbf{k}_w,\omega/c)$  is a \emph{necessary condition} required by Einstein's special relativity.  Since $K^{\mu}=(\mathbf{k}_w,\omega/c)$  is a Lorentz four-vector, $K^{\mu}K_{\mu}$ must be a Lorentz invariant, leading to $\omega^2(1-n^2)=$ Lorentz invariant.  However Partanen and coworkers claim that the energy $E_{\textrm{MP}}=\hbar\omega+(\delta m)c^2=n^2\hbar\omega$ and momentum  $p_{_{\mathrm{MP}}}=n\hbar\omega/c$  for their MP quasiparticle also constitute a Lorentz four-vector \cite{r1}, leading to $(n\omega)^2(n^2-1)=$ Lorentz invariant.  From $\omega^2(1-n^2)=$ Lorentz invariant and $(n\omega)^2(n^2-1)=$ Lorentz invariant, it follows that both the frequency $\omega$  and the refractive index $n$ are Lorantz invariants for $n\ne 1$, which means that there is no Doppler effect.  Obviously, this result is not physical.  Thus we conclude that Partanen-coworkers physical model is not compatible with Einstein's special relativity.
} and $\mathbf{k}_w$  is the wave vector.   $\hbar K^{\mu}$ indeed does not include ``the transferred mass $\delta m$'', but it is four-vector covariant \cite{r5}, never leading to any mathematical problems; thus Partanen-coworkers criticism is not pertinent either. 

For a plane light wave propagating in a uniform medium, \emph{microscopically} there is a complicated interaction between photons and matter through atomic absorption and re-emission.  However \emph{macroscopically}, this interaction can be simply described by Minkowski (space-like) four-momentum  $\hbar K^{\mu}$; from this view of point, Minkowski photon is a kind of ``quasi-photon'', with an imaginary ``rest mass'' if defined by $(\hbar K^{\mu}\hbar K_{\mu})^{1/2}/c$ \cite{r3}, which has no physical meaning.  However in the Einstein's energy-mass equivalence equation, the definition of mass for a photon, including the rest mass, is the energy-associated mass, given by $\hbar\omega/c^2$ \cite{r7}.\footnote{
In Ref.\,\cite{r7},  by analyzing a special radiation process based on the principle of relativity, Einstein draws a conclusion that ``If a body gives off the energy $E$ in the form of radiation, its mass diminishes by $E/c^2$.''  Thus according to Einstein, with energy conservation law taken into account, the energy-associated mass for a photon is supposed to be $\hbar\omega/c^2$. 
} Thus the introduction of quasi-photon has no contradiction against the traditional theory. (Note: In free space, the rest mass of photon has no physical meaning, because there is no photon-rest frame in such a case.)

It should be emphasized that the Lorentz covariant wave four-vector $K^{\mu}$ is a strict result from special relativity, and its covariance insures that under Lorentz transformation of  $K^{\mu}=(\mathbf{k}_w,\omega/c)$,  $\mathbf{k}_w$ and $\omega$  keep the same definitions and mathematical forms in all inertial frames, while Partanen-coworkers MP quasiparticle \cite{r1} is not compatible with $K^{\mu}$, as shown in footnote 3.  In other words, the physical model used in Partanen-coworkers theory is not consistent with the principle of relativity. 

In my theory, the definitions of all physical quantities, including the refractive index $n$, are the same in all inertial frames.  If Minkowski photon momentum and energy in the medium-rest frame are given by 
\begin{equation} 
\mathbf{p}=\hbar\mathbf{k}_w,~~~~~~~~~~E=\hbar\omega,						
\label{eq5}
\end{equation}  
where $\mathbf{k}_w=\mathbf{\hat{n}}|n\omega/c|$  with $\mathbf{\hat{n}}=\mathbf{k}_w/|\mathbf{k}_w|$  the unit wave vector, then observed in the frame moving at a velocity of $\mathbf{v}$  with respect to the medium-rest frame, the momentum and energy are given by \\
\begin{equation} 
\mathbf{p}'=\hbar\mathbf{k}'_w,~~~~~~~~~~E'=\hbar\omega',						
\label{eq6}
\end{equation} 
where $\mathbf{k}'_w=\mathbf{\hat{n}}'|n'\omega'/c|$  with $\mathbf{\hat{n}}'=\mathbf{k}'_w/|\mathbf{k}'_w|$ the unit wave vector.  Note that Eq.\,(\ref{eq5}) and Eq.\,(\ref{eq6}) have completely the same forms, characterizing the covariance of the \emph{momentum and energy} of light.  The plane wave propagates at the phase velocity  $\mathbf{v}_{ph}=\mathbf{\hat{n}}(\omega/|\omega|)(c/n)$, which holds in all inertial frames, characterizing the covariance of the \emph{propagation direction and speed} of light. These basic properties of light are required by the principle of relativity: Physical laws are symmetric for all inertial frames, and no one is preferred for descriptions of physical phenomena. It is worthwhile to point out that Minkowski photon momentum satisfies Newton law \cite{r3}. \\
\indent When observed in the \emph{photon-rest} frame, namely the frame moving at a velocity of $\mathbf{v}=\mathbf{\hat{n}}(c/n)$  with respect to the medium-rest frame, the momentum and energy become 
\begin{equation} 
\mathbf{p}'=\mathbf{\hat{n}}'\frac{\hbar\omega}{c}\sqrt{n^2-1},~~~~~~~~E'=0,						
\label{eq7}
\end{equation} 
where $n>1$ is assumed. In such a case, $\omega'=0$  but  $n'=\infty$ \cite{r5}, resulting in finite $n'\omega'=\omega(n^2-1)^{1/2}$.  Thus the Minkowski photon is stopped there ($c/n'=0$), with a zero energy ($E'=\hbar\omega'=0$) and a finite nonzero momentum of  $|\mathbf{p}'|=(n^2-1)^{1/2}\hbar\omega/c$. \\
\indent Hau and coworkers experimentally demonstrated that light can either be stopped or slowed down significantly to a speed of 17 meters per second when it goes into an ultracold atom gas in the Bose-Einstein condensate \cite{r11}.\footnote{
In this experimental report, Hau and coworkers argue that the slowdown of light pulses is caused by the strong dispersion of medium which results in a low group velocity.  However the group velocity is not an observable quantity \emph{in general}, because the physical implication of the definition of group velocity itself is ambiguous \cite{r15}. But for a monochromatic plane wave or a single photon, the dispersion will not have effect.  The plane wave is not practical, but a single photon can be realized in current technologies.  Thus Hau-coworkers argument could be experimentally verified by use of a single photon.
}  Suppose that macroscopically this atom gas can be treated as a uniform medium with refractive index $n$ when a \emph{monochromatic} plane light wave propagates through.  In such a case, from my theory, Minkowski photon would have a very small or zero energy, namely $E=\hbar\omega\rightarrow 0$, because the momentum $|\mathbf{p}|=n\hbar\omega/c$ is finite when  $n\rightarrow\infty$.  Thus the result from my theory makes sense.

What is the difference between my theory and Minkowski's theory?  Three years after Einstein developed the special theory of relativity and showed that wave vector $\mathbf{k}_w$ and frequency $\omega$ constitute a wave four-vector for a plane wave in \emph{free space} \cite{r8}, Minkowski applied Einstein's principle of relativity to a \emph{moving medium}, arguing for invariance of Maxwell equations, namely Maxwell equations have the same form no matter whether the medium is moving uniformly or at rest \cite{r12}.  This invariance is the foundation of Minkowski's electrodynamics of moving media, and it is widely accepted in the community \cite{r13}.  However, I firmly argue that to examine the covariance of light propagation in a medium, the definitions of all physical quantities, including the refractive index, \emph{also} must be the same in all inertial frames. In addition, I have shown 
that within the principle-of-relativity frame, the momentum and energy of a photon in a medium must constitute a Lorentz four-vector \emph{to satisfy global momentum-energy conservation law} 
in the Einstein-box thought experiment \cite{r3}, and \emph{to satisfy Fermat's principle} for a plane wave in a uniform medium \cite{r5}.
  Thus compared with Minkowski's theory, my theory takes into account not only (i) the invariance of Maxwell equations, but also (ii) the invariance of definitions of physical quantities, and (iii) the four-vector covariance of momentum and energy of light \cite{r5}.  Note that all conditions (i), (ii), and (iii) are required by the principle of relativity.

Obviously, conditions (i) and (ii) have already ensured the covariance of physical laws, namely all the equations describing the behavior of light in a medium have the same forms in all inertial frames.  However without condition (iii), the definition of light momentum is indefinite.  For example, under (i) and (ii) without (iii), both Abraham formulation $\mathbf{g}_{_A}=\mathbf{E}\times\mathbf{H}/c^2$ and Minkowski formulation  $\mathbf{g}_{_M}=\mathbf{D}\times\mathbf{B}$ have the equal right to be the light momentum density vector, because they all have the same forms in all inertial frames, namely they are all covariant together with Maxwell equations under Lorentz transformations of field-strength \emph{four-tensors} (confer footnote 7).  However when condition (iii) is considered, only $\mathbf{g}_{_M}=\mathbf{D}\times\mathbf{B}$ is the qualified light momentum vector, because $\mathbf{g}_{_M}=\mathbf{D}\times\mathbf{B}$ is also \emph{four-vector covariant}, as the space component of the four-vector  $\hbar K^{\mu}=N_p^{-1}(\mathbf{g}_{_M},W_{em}/c)$, as shown in Ref.\,\cite{r5}, while $\mathbf{g}_{_A}=\mathbf{E}\times\mathbf{H}/c^2$ is only ``four-tensor covariant'', instead of also ``four-vector covariant''; in other words, although $\mathbf{g}_{_A}=\mathbf{E}\times\mathbf{H}/c^2$ is covariant, it cannot be used to form a momentum-energy four-vector, as shown in Ref.\,\cite{r3}. Thus condition (iii) is indispensable for a complete covariant theory of light propagation. 

It might be interesting to point out that the invariance of Maxwell equations in Minkowski's electrodynamics of \emph{moving media} is  realized based on the condition that all EM fields constitute two field-strength tensors \cite{r12,r14}.  From this it follows that the wave phase $\Psi(\mathbf{x},t)=(\omega t-\mathbf{k}_w\cdot\mathbf{x})$ must be a Lorentz invariant, and the frequency $\omega$  and the wave vector $\mathbf{k}_w$  must constitute a Lorentz four-vector $K^{\mu}=(\mathbf{k}_w,\omega/c)$ \cite{r5,r10}. However this fundamental physical result from the Minkowski's electrodynamics is usually ignored in the analysis of Abraham-Minkowski debate, such as in the work by Partanen and coworkers \cite{r1}, although Einstein has already made a pioneer demonstration in his 1905 paper for a plane wave in \emph{free space} \cite{r8}. 

\section{Conclusions and remarks}
\label{s5}
In this paper, two new concepts (Minkowski quasi-photon and invariance of physical definitions) have been introduced to elucidate the theory developed in my previous work \cite{r5}, and to clarify the criticisms by Partanen and coworkers \cite{r1}.

Specifically speaking, the criterion of Lorentz four-vector covariance set up by Partanen and coworkers is not mathematically correct, and the physical model they constructed is not compatible with Einstein's special relativity, as shown in footnote 3.  In contrast, my theory is a self-consistent covariant theory of light propagation in a medium \cite{r5}; Minkowski quasi-photon appearing in my theory comes from the wave four-vector, which is a well-known strict result from Einstein's special relativity, just like the time dilation, Lorentz contraction, and relativity of simultaneity. According to Einstein's energy-mass equivalence equation, the mass of a photon is always defined by  $\hbar\omega/c^2$ \cite{r7}, and the introduction of quasi-photon is compatible with the relativity. Thus it is not appropriate for Partanen-coworkers criticism that my theory ``neglects the transferred mass  $\delta m$, thus leading to mathematical problems'', and ``the neglectance of the transferred mass  $\delta m$ ... in turn leads to complicated mathematics without providing transparent and physically insightful covariant theory of light'' \cite{r1}. 

It should be emphasized that my theory \cite{r5} takes into account not only (i) the invariance of Maxwell equations argued by Minkowski \cite{r12}, but also (ii) the invariance of physical definitions, which is first clearly formulated in this paper, and (iii) the four-vector covariance of momentum and energy of light \cite{r3}.  All conditions (i), (ii), and (iii) are required by the principle of relativity, and without (iii), the definition of light momentum in a medium is indefinite. 

It is also worthwhile to point out that above conditions (i) and (ii) are to fulfill the symmetry of Maxwell EM theory required by the principle of relativity, while condition (iii) is to fulfill the symmetries of momentum-energy conservation law \cite{r3} and Fermat's principle \cite{r5}, which are additional basic postulates in physics, independent of Maxwell EM theory \cite{r15}.\footnote{
In Ref.\,\cite{r15}, it is shown that Fermat's principle is an independent postulate, and it \emph{cannot} be derived from Maxwell EM theory, although there is a formulation of Fermat's principle in the Maxwell-equation frame.  The original form of Fermat's principle states that \emph{Nature always acts by the shortest course}.  When applied to optics, this principle requires that \emph{light take the path of least time}.  In optics of light rays (geometric optics) set up from Maxwell equations, this principle is specifically expressed as: An actual light ray makes its optical length $\int n\mathrm{d}s$ the minimum.  Light rays, refractive index, light speed ... are all defined in the Maxwell-equation frame, but the conclusion ``an actual light ray makes its optical length $\int n\mathrm{d}s$ the minimum'' comes from the Fermat's principle.  In other words, Fermat's principle itself is not included in Maxwell equations.  It is also shown that the energy conservation law (instead of energy conservation equation or Poynting theorem) is not included in but is consistent with Maxwell EM theory, and it is also an independent postulate in physics.
}  Both Abraham momentum $\mathbf{g}_{_A}=\mathbf{E}\times\mathbf{H}/c^2$ and Minkowski momentum $\mathbf{g}_{_M}=\mathbf{D}\times\mathbf{B}$ satisfy the Maxwell-EM-theory symmetry, but only the Minkowski momentum satisfies all the above three symmetries --- explaining why it is controversial to define the momentum of light only in the Maxwell-EM-theory fame.  

Macroscopic Maxwell equations are the foundation of classical electrodynamics \cite{r10,r14}, while Einstein's light-quantum hypothesis is the basis of Bohr frequency condition in atomic transitions that cause the absorption or emission of photons \cite{r16,r17}.  Macroscopic Maxwell equations combined with Einstein's light-quantum hypothesis lead to a self-consistent Minkowski photon four-momentum $\hbar K^{\mu}=N_p^{-1}(\mathbf{g}_{_M},W_{em}/c)$  for a plane light wave \cite{r5}, and thus this four-momentum macroscopically describes the light-matter interactions through atomic absorption and re-emission.  On the other hand, a material medium is made up of massive particles, and all individual massive particles always have their own momentum-energy four-vectors no matter whether there are any interactions \cite{r3}. From this perspective, when a closed system of light-matter interactions is divided into an EM subsystem and a material subsystem \cite{r18,r19}, \emph{the momentums and energies of the two subsystems constitute Lorentz four-vectors, respectively}, while the Minkowski photon four-momentum $\hbar K^{\mu}$ denotes the macroscopic average of the properties of photons absorbed and re-emitted by the material subsystem.  \\
\indent In terms of the relations between EM fields and the materials that generate or support the fields, the EM fields can be divided into two types: One is radiation field and the other is non-radiation field.  The radiation field can exist independently of the material that generates it, such as the EM field carried by a light pulse, of which all individual photons have their own momentum-energy four-vectors.  The non-radiation field cannot exist separately from the material that supports it, such as the electrostatic field in vacuum, of which the momentum and energy cannot form a momentum-energy four-vector \cite{r20}.  From this, \emph{a non-radiation field is owned by the material that supports the non-radiation field, and its momentum and energy are, respectively, included in the momentum and energy of the material subsystem} (instead of the EM subsystem). To better understand this, let us take a look at the following example.  \\
\indent Consider a closed physical system in free space, where there is only one free electron at rest in the laboratory frame $\Sigma$.   In such a case, the electrostatic field produced by the rest electron is a self-field (non-radiation field), and it cannot exist separately; thus its energy is a part of the electron's rest energy $m_ec^2$ (= 0.511 MeV), where $m_e$ is the electron's rest mass, although we don't know how much the part is --- depending on what physical model of the electron's field distribution is assumed.  Thus the EM subsystem is empty, and the material subsystem is equal to the whole closed system.  Observed in a frame $\Sigma'$  that moves uniformly with respect to the laboratory frame $\Sigma$ at a velocity of $\mathbf{v}$, the electron has a momentum-energy four-vector, given by $P'^{\mu}=(\mathbf{p}',E'/c)$, where $\mathbf{p}'=\gamma\hspace{0.4mm}' m_e\mathbf{v}'$ and $E'=\gamma\hspace{0.4mm}' m_ec^2$, with $\mathbf{v}'=-\mathbf{v}$ and $\gamma\hspace{0.4mm}'=(1-\mathbf{v}'^2/c^2)^{-1/2}$.  The electron's momentum $\mathbf{p}'=\gamma\hspace{0.4mm}' m_e\mathbf{v}'$  and energy $E'=\gamma\hspace{0.4mm}' m_ec^2$  include the momentum and energy of the self-EM field --- non-radiation field --- carried by the moving electron, while the electron's kinetic energy $(\gamma\hspace{0.4mm}'-1)m_ec^2$ \cite{r8} includes the kinetic energy of the self-EM field.  Therefore,  unlike a radiation field, the non-radiation field carried by a charged particle is always associated with the particle itself and cannot be separated; thus the non-radiation field, as a part of the particle, belongs to the material subsystem, instead of the EM subsystem.  One might be able to measure the momentum and energy of a moving electron in free space, but one cannot identify how much the EM counterparts occupy; in other words, the momentum and energy of a non-radiation field are not measurable experimentally. 

The theoretical analysis of Compton photon-electron scattering experiment \cite{r21} is a typical example in which the non-radiation field is treated to be owned by the material subsystem.  According to Einstein's light-quantum hypothesis \cite{r22} and energy-mass equivalence equation \cite{r7}, Compton wrote down the photon's momentum in vacuum, given by $(\hbar\omega/c^2)\times c=\hbar\omega/c$  (momentum = mass multiplied velocity), which is completely consistent with later-published de-Broglie hypothesis \cite{r23}, while the electron's momentum and energy are given by  $\gamma m_e\mathbf{v}$  and  $\gamma m_ec^2$, respectively, with $\mathbf{v}$  the electron's velocity and $\gamma=(1-\mathbf{v}^2/c^2)^{-1/2}$.  Before and after the scattering, the momentum and energy are conservative, and thus Compton obtained a formula of photon-electron scattering, which is in very good agreement with experimental results.  In his derivations, Compton \emph{never} considered the momentum and energy of the (non-radiation) self-EM field carried by the electron.  Thus the Compton experiment is an apparent support to the conclusion obtained above:  a non-radiation field is the component of the material subsystem, instead of the EM subsystem. \\
\indent Finally, I would like to make some comments on the symmetry and covariance of physical laws, and the invariance of the definitions of all physical quantities. \\
  \indent The role of symmetry in fundamental physics is well reviewed \cite{r24}.  Symmetry and covariance usually refer to the mathematical forms of physical laws.  For example, according to the principle of relativity, Einstein recognized that all inertial frames are equivalent, and thus Maxwell equations must be symmetric or invariant in form, although the values of corresponding EM field quantities between different inertial frames may be different. To keep Maxwell equations the same in form (symmetry), the EM fields must follow Lorentz transformations of field-strength tensors; thus Maxwell equations are also said to be \emph{covariant} under Lorentz transformations \cite[p.553]{r14}, which means: (i) field-strength tensors follow Lorentz transformations, and (ii) Maxwell equations keep invariant in form --- \emph{co-variant}.  To put it simply, we have: principle of relativity $\Rightarrow$ equivalence of inertial frames  $\Rightarrow$ symmetry of Maxwell equations  $\Rightarrow$ covariance of Maxwell equations under Lorentz transformations. \\
\indent Why should the definitions of all physical quantities be invariant?  The principle of relativity states that the results of experiments carried through in isolated laboratories in inertial frames are independent from the orientation and the velocity of the laboratory \cite{r25}; in other words, all inertial frames are equivalent, and no one is preferred for descriptions of physical phenomena.  Thus the Maxwell EM theory must be symmetric; and all EM principles are the same in every inertial frame.  A specific EM principle is expressed by incorporating physical quantities; thus the definitions of these physical quantities must be the same to keep the principle invariant. \\
\indent We also can obtain this conclusion directly from the equivalence of inertial frames, which is an elegant proof for the invariance of physical definitions. Suppose that $\Sigma$  and $\Sigma'$  are any two inertial frames moving with respect to each other.  If frame $\Sigma$  defines a physical quantity in this way while frame  $\Sigma'$ defines the same physical quantity in that way, and they both claim to be flaw-free, then who is correct?  The only possible resolution of this debate is to use the same definition, because $\Sigma$ and  $\Sigma'$  become the same frame when the relative velocity between them goes to zero --- \emph{the continuity of physical definitions}, just like the continuity of Lorentz transformations argued by Einstein \cite{r8}. \\
\indent In fact, the symmetry of Maxwell equations has already suggested the invariance of physical definitions. As we know, in the Maxwell equations $\nabla\times\mathbf{E}=-\partial\mathbf{B}/\partial t$, $\nabla\cdot\mathbf{B}=0$,  $\nabla\times\mathbf{H}=\mathbf{J}+\partial\mathbf{D}/\partial t$, and $\nabla\cdot\mathbf{D}=\rho$, all the field quantities $\mathbf{E}$, $\mathbf{B}$, $\mathbf{H}$, and $\mathbf{D}$, and current density $\mathbf{J}$ and charge density $\rho$  have the same definitions in all inertial frames.\footnote{
``Lorentz covariance'' itself means that under Lorentz transformations, something required by the principle of relativity keeps invariant.  The principle of relativity argues that all the inertial frames of reference are equivalent, while the invariance of physical definitions is a reflection of the equivalence.  Thus the Lorentz covariance can be defined as follows.  A physical quantity is said to be Lorentz covariant if its physical definition keeps the same in all inertial frames under Lorentz transformations; for example, the refractive index $n$ for a plane wave is Lorentz covariant because it has the same definition in all inertial frames \cite{r5}. A tensor is said to be Lorentz covariant if all the elements of the tensor keep the same physical definitions under Lorentz transformations.  For example, the wave phase $\Psi=(\omega t-\mathbf{k}_w\cdot\mathbf{x})=g_{\mu\nu}K^{\mu}X^{\nu}$ is covariant zeroth-rank tensor, and the time-space four-vector $X^{\mu}=(\mathbf{x},ct)$ is covariant first-rank tensor, because their elements, $(\omega t-\mathbf{k}_w\cdot\mathbf{x})$ for $\Psi$, and $\mathbf{x}$ and $ct$ for $X^{\mu}$, have the same physical definitions in all inertial frames.  As shown in the textbook by Jackson \cite[p.\,557]{r14}, 3D-form Maxwell equations $[\nabla\times\mathbf{H}-\partial(c\mathbf{D})/\partial(ct),\nabla\cdot(c\mathbf{D})]=(\mathbf{J},c\rho)$ and $[\nabla\times\mathbf{E}-\partial(-c\mathbf{B})/\partial(ct),\nabla\cdot(-c\mathbf{B})]=(\mathbf{0},0)$ can be written in tensor form, given by $\partial_{\mu}G^{\mu\nu}(\mathbf{D},\mathbf{H})=J^{\nu}$ and $\partial_{\mu}\mathscr{F}^{\mu\nu}(\mathbf{B},\mathbf{E})=0$, with $G^{\mu\nu}$ being the field-strength four-tensor, $\mathscr{F}^{\mu\nu}$ being the dual field-strength four-tensor of $F^{\mu\nu}=\partial^{\mu}A^{\nu}-\partial^{\nu}A^{\mu}$, and $J^{\mu}=(\mathbf{J},c\rho)$ being the four-current density.  To keep Maxwell equations invariant in form (namely the symmetry required by the principle of relativity), $G^{\mu\nu}$ and $\mathscr{F}^{\mu\nu}$ must be assumed to be Lorentz covariant, namely (i) $G^{\mu\nu}$ and $\mathscr{F}^{\mu\nu}$ follow Lorentz transformations, and (ii) the elements of $G^{\mu\nu}$ and $\mathscr{F}^{\mu\nu}$ have the same physical definitions in all inertial frames.  If the definitions of the elements were not the same, then the physical implications of Maxwell equations would not be the same, which is not consistent with the principle of relativity.  [Note that in such a case, the covariance of $J^{\mu}$ is automatically fulfilled, because $\partial_{\mu}=(\partial/\partial\mathbf{x},\partial/\partial ct)$ is Lorentz covariant four-vector operator, namely $\partial_{\mu}=(\partial/\partial\mathbf{x},\partial/\partial ct)$ follows Lorentz transformation with its elements $\partial/\partial\mathbf{x}$ and $\partial/\partial ct$ have the same definition in all inertial frames.]  Thus the covariance of $G^{\mu\nu}$ and $\mathscr{F}^{\mu\nu}$  provides a guarantee of the symmetry of Maxwell equations between inertial frames; in other words, this covariance offers an assurance of the invariance in form of Maxwell equations under Lorentz transformations, which is usually referred to as Lorentz covariance of Maxwell equations \cite[p.553]{r14}.   The tensor-form Maxwell equations $\partial_{\mu}G^{\mu\nu}(\mathbf{D},\mathbf{H})=J^{\nu}$ and $\partial_{\mu}\mathscr{F}^{\mu\nu}(\mathbf{B},\mathbf{E})=0$ are also said to be Lorentz covariant because they are expressed through covariant Lorentz tensors.  The 3D-form and tensor-form of Maxwell equations are equivalent when the Lorentz covariance of $G^{\mu\nu}$ and $\mathscr{F}^{\mu\nu}$ is satisfied, and all the eight 3D-Maxwell (component) equations are the eight elements of the two tensor-Maxwell equations.  As we know, the Minkowski EM stress-energy tensor is constructed from the two field-strength tensors $G^{\mu\nu}$ and $F^{\mu\nu}$, given by $T^{\mu\nu}=g^{\mu\sigma}G_{\sigma\lambda}F^{\lambda\nu}+0.25g^{\mu\nu}G_{\sigma\lambda}F^{\sigma\lambda}$, with $c\mathbf{g}_{_A}=\mathbf{E}\times\mathbf{H}/c$ and $c\mathbf{g}_{_M}=c\mathbf{D}\times\mathbf{B}$ as its elements \cite{r20}; thus the Lorentz covariance of $T^{\mu\nu}$ ensures that Abraham momentum $\mathbf{g}_{_A}=\mathbf{E}\times\mathbf{H}/c^2$ and Minkowski momentum $\mathbf{g}_{_M}=\mathbf{D}\times\mathbf{B}$ have the same definitions and mathematical forms in all inertial frames, namely they are both covariant together with Maxwell equations from one inertial frame to another. According to the definition of Lorentz covariance of a tensor given above, Partanen and coworkers did not provide a proof of the covariance of four-momentum in their MP quasiparticle model, because the refractive index $n$ appearing in $E_{\mathrm{MP}}=n^2\hbar\omega$ and $p_{_{\mathrm{MP}}}=n\hbar\omega/c$ is not defined in a general moving frame \cite{r1} and one cannot judge whether the elements $E_{\mathrm{MP}}$ and $p_{_{\mathrm{MP}}}$ of the four-momentum have the same physical definitions in all inertial frames.
}  Maxwell equations are physical laws describing macroscopic EM phenomena, and they are expressed by the physical quantities that have the same definitions in all inertial frames; this is a basic property assigned by the principle of relativity.  All physical laws have the same right to share this property with Maxwell equations.  Specifically, the refractive index $n$, which is a physical quantity to describe the covariance of the propagation speed $(c/n)$ of light in a medium, is supposed to have the same definition in all inertial frames, as shown in Ref.\,\cite{r5}. \\
\indent From above we can see that the invariance of physical definitions is a generalization of the symmetry of physical laws within the frame of relativity principle. \\ 








\newpage
\onecolumn

\begin{figure} 
\includegraphics[trim=1.0in 1.0in 1.0in 1.0in, clip=true,scale=1.0]{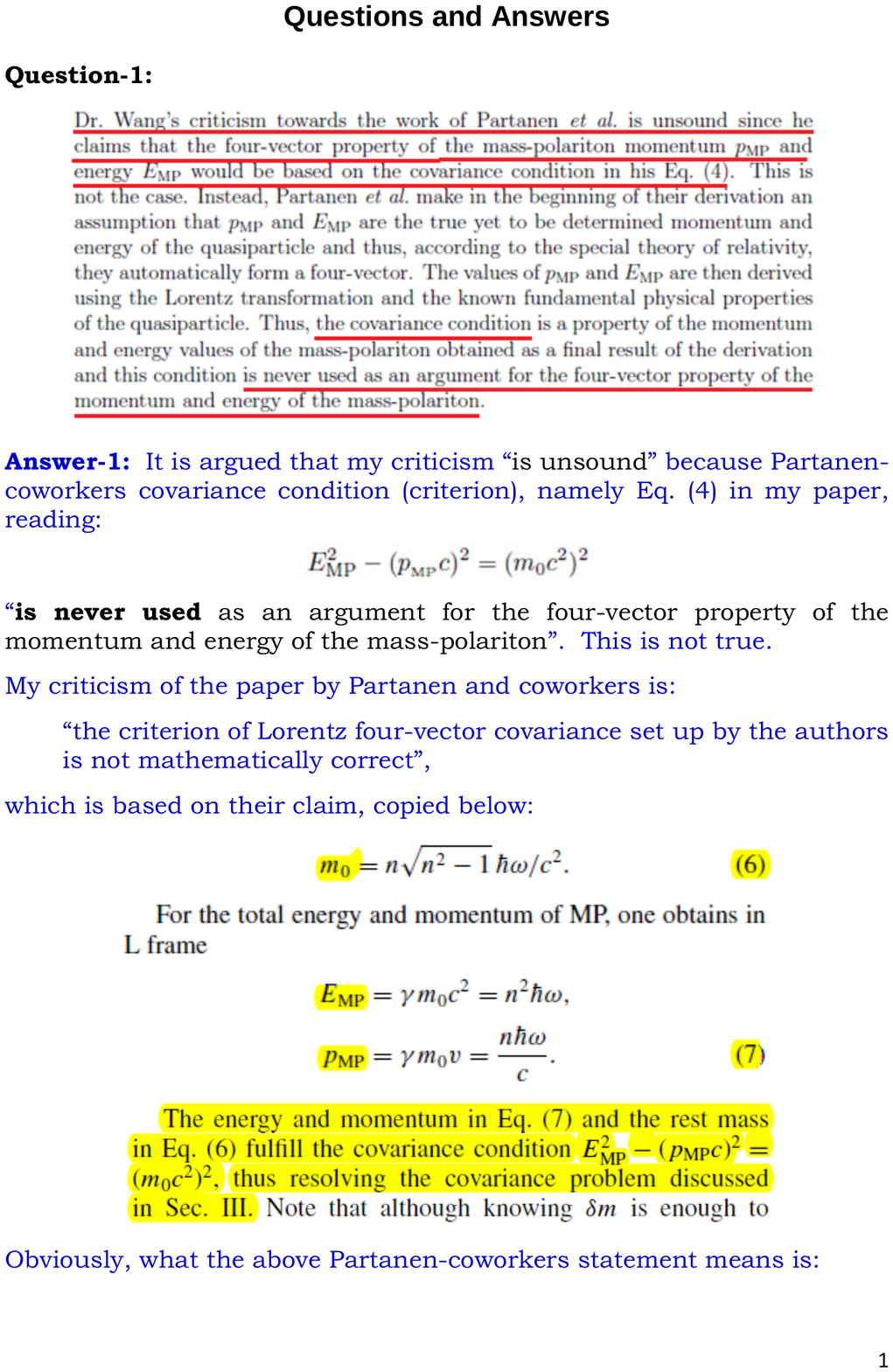}
\label{figM1}
\end{figure} 

\begin{figure} 
\includegraphics[trim=1.0in 1.0in 1.0in 1.0in, clip=true,scale=1.0]{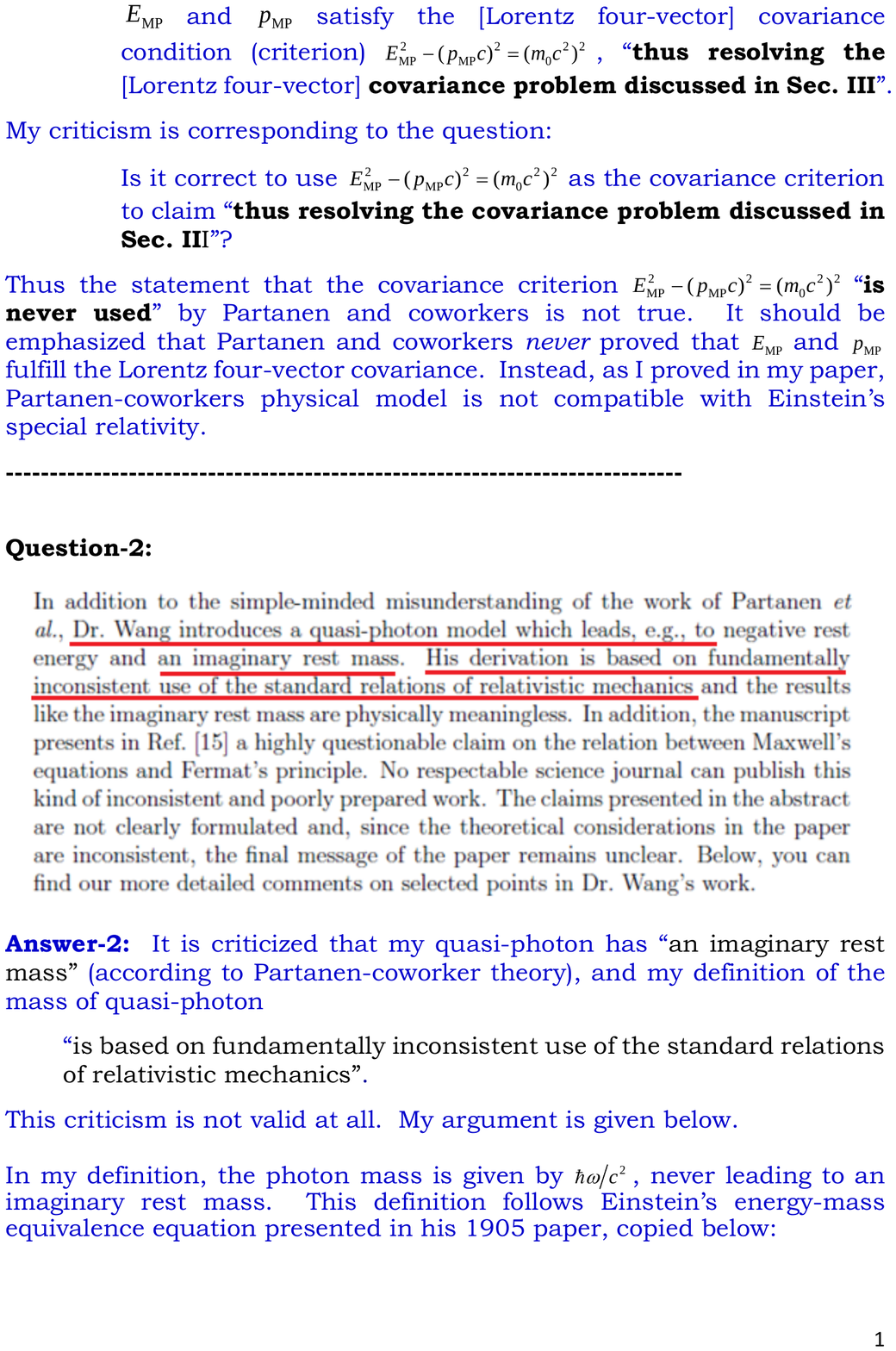}
\label{figM2}
\end{figure} 

\begin{figure} 
\includegraphics[trim=1.0in 1.0in 1.0in 1.0in, clip=true,scale=1.0]{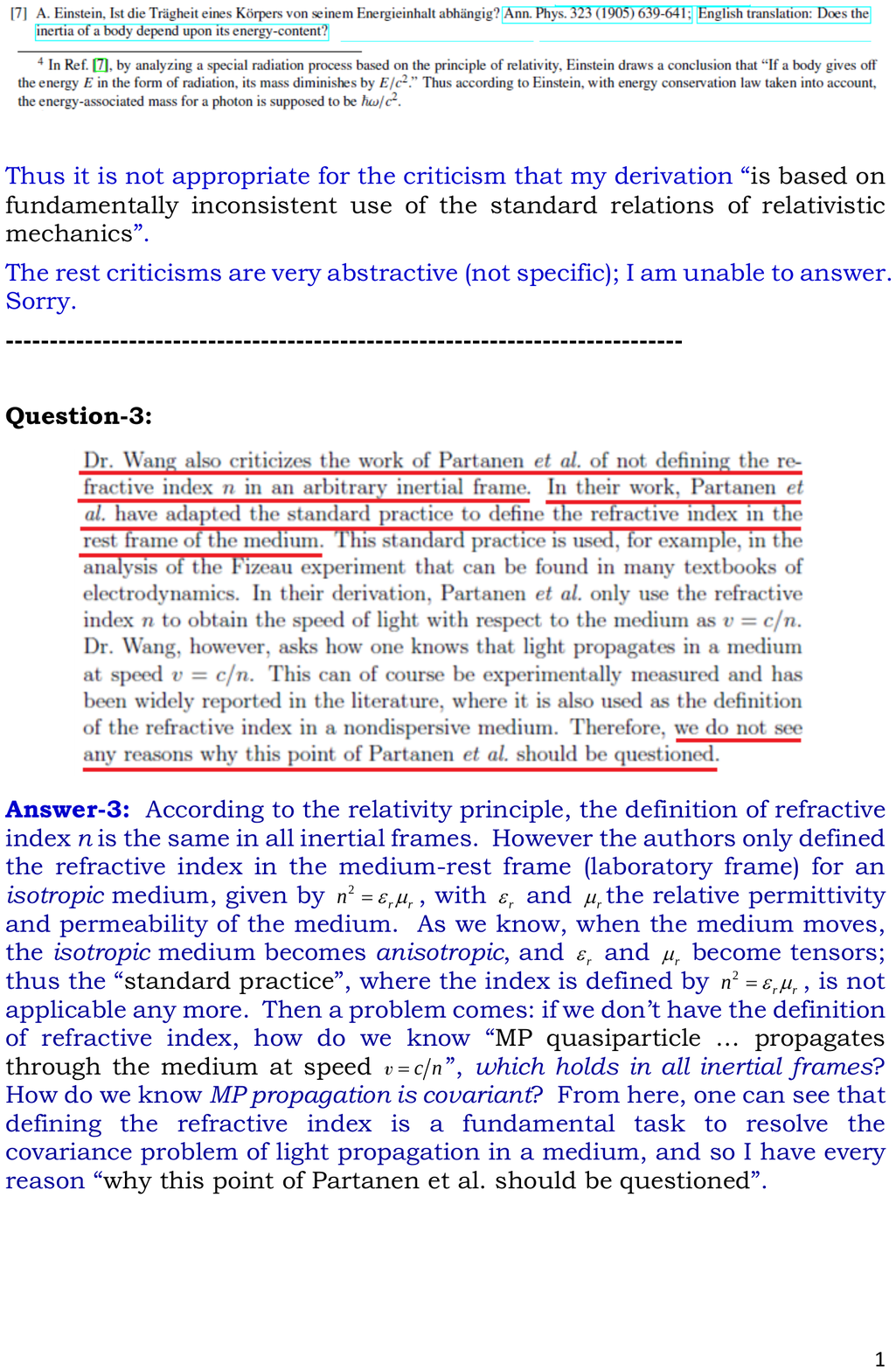}
\label{figM3}
\end{figure} 

\begin{figure} 
\includegraphics[trim=1.0in 1.0in 1.0in 1.0in, clip=true,scale=1.0]{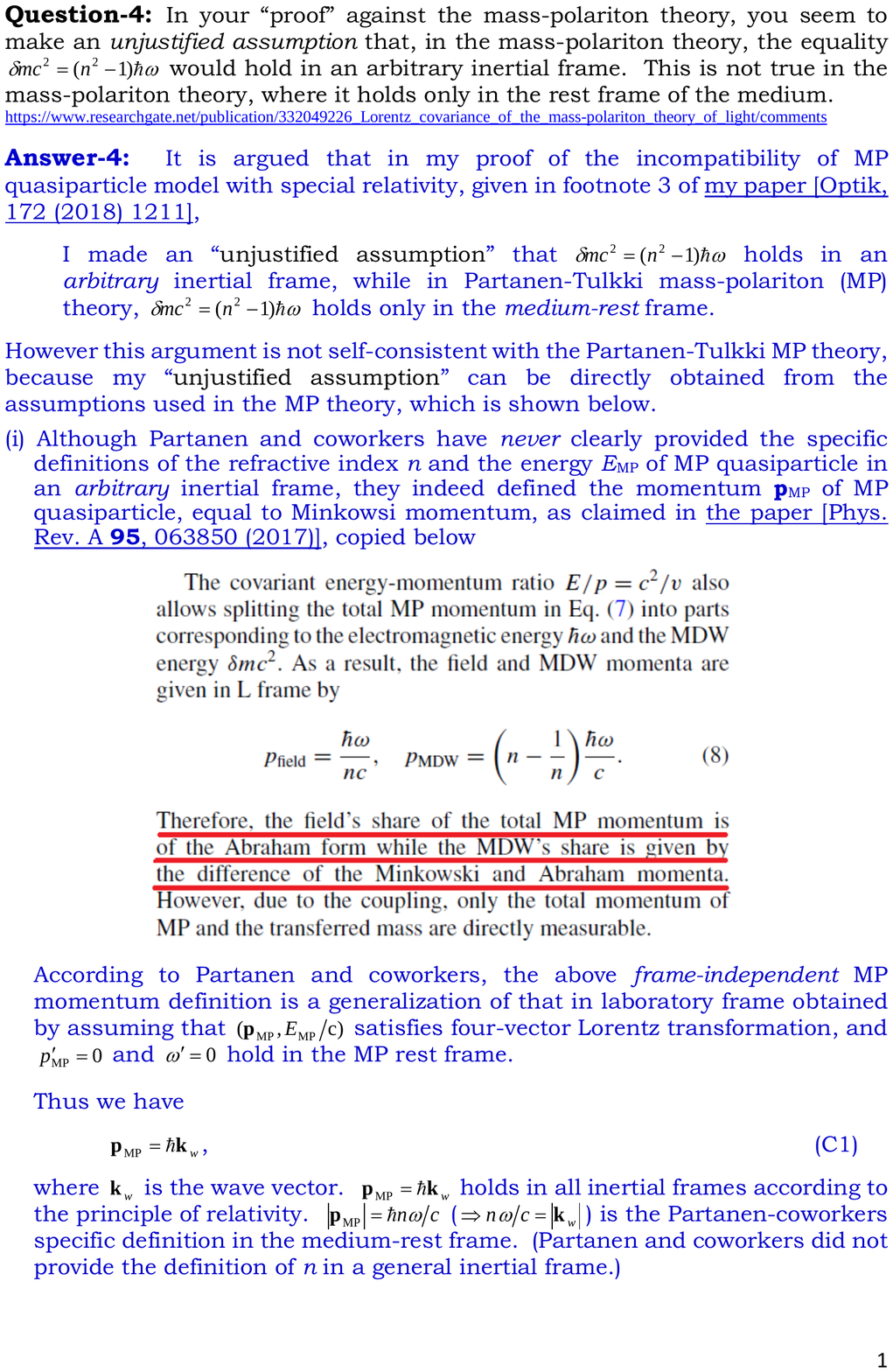}
\label{figM4}
\end{figure} 

\begin{figure} 
\includegraphics[trim=1.0in 1.0in 1.0in 1.0in, clip=true,scale=1.0]{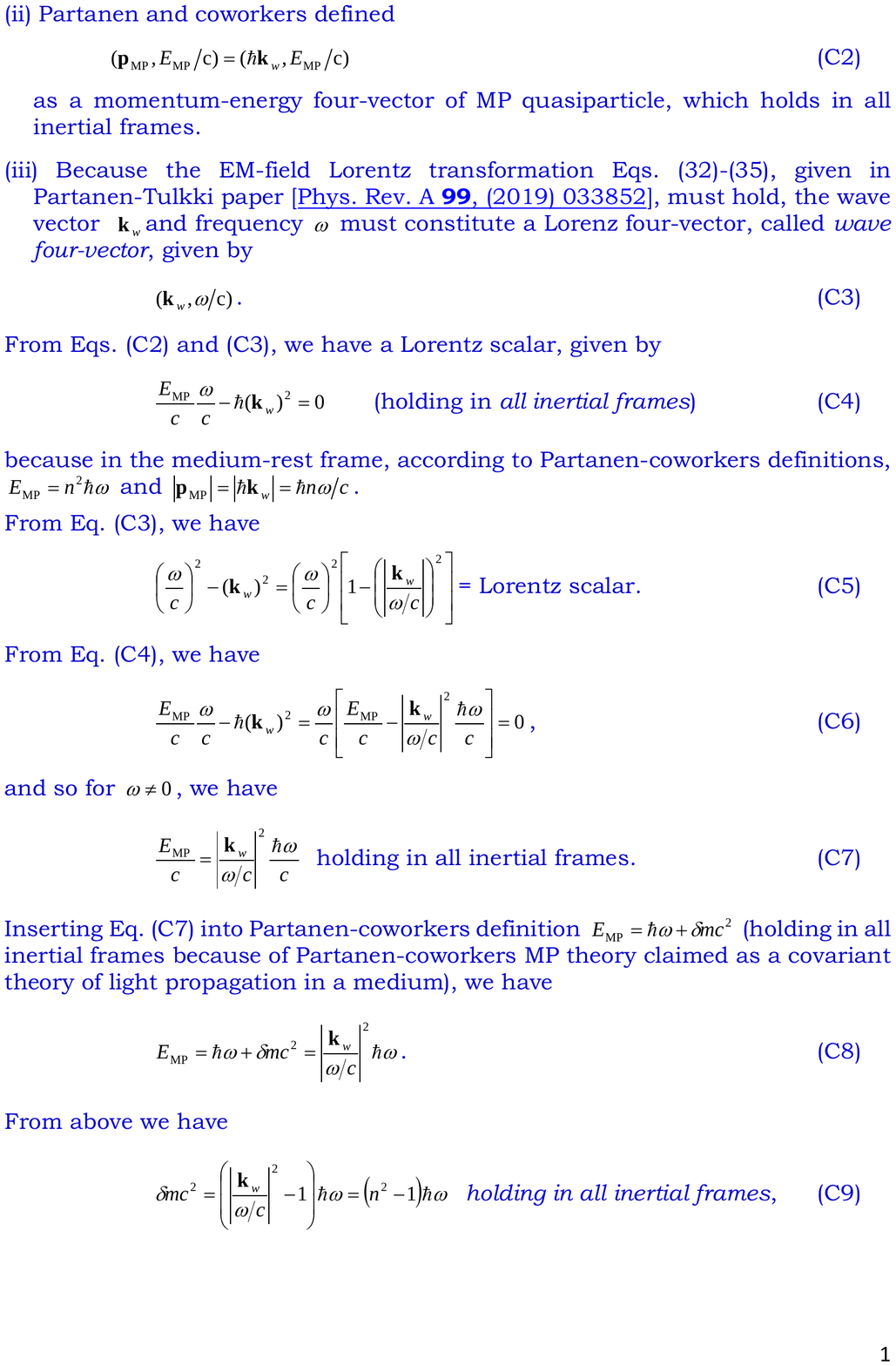}
\label{figM5}
\end{figure} 

\begin{figure} 
\includegraphics[trim=1.0in 1.0in 1.0in 1.0in, clip=true,scale=1.0]{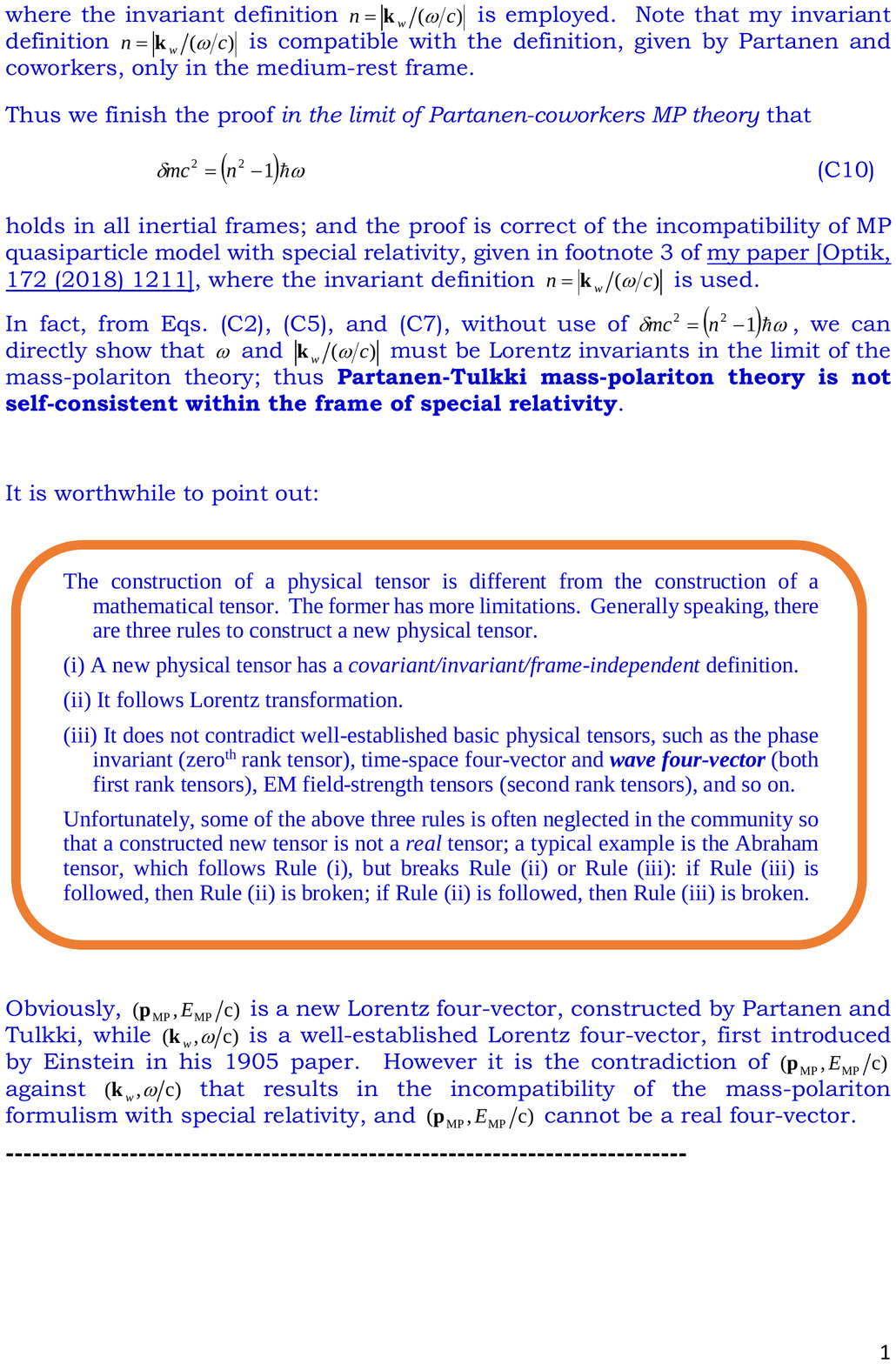}
\label{figM6}
\end{figure} 

\begin{figure} 
\includegraphics[trim=1.0in 1.0in 1.0in 1.0in, clip=true,scale=1.0]{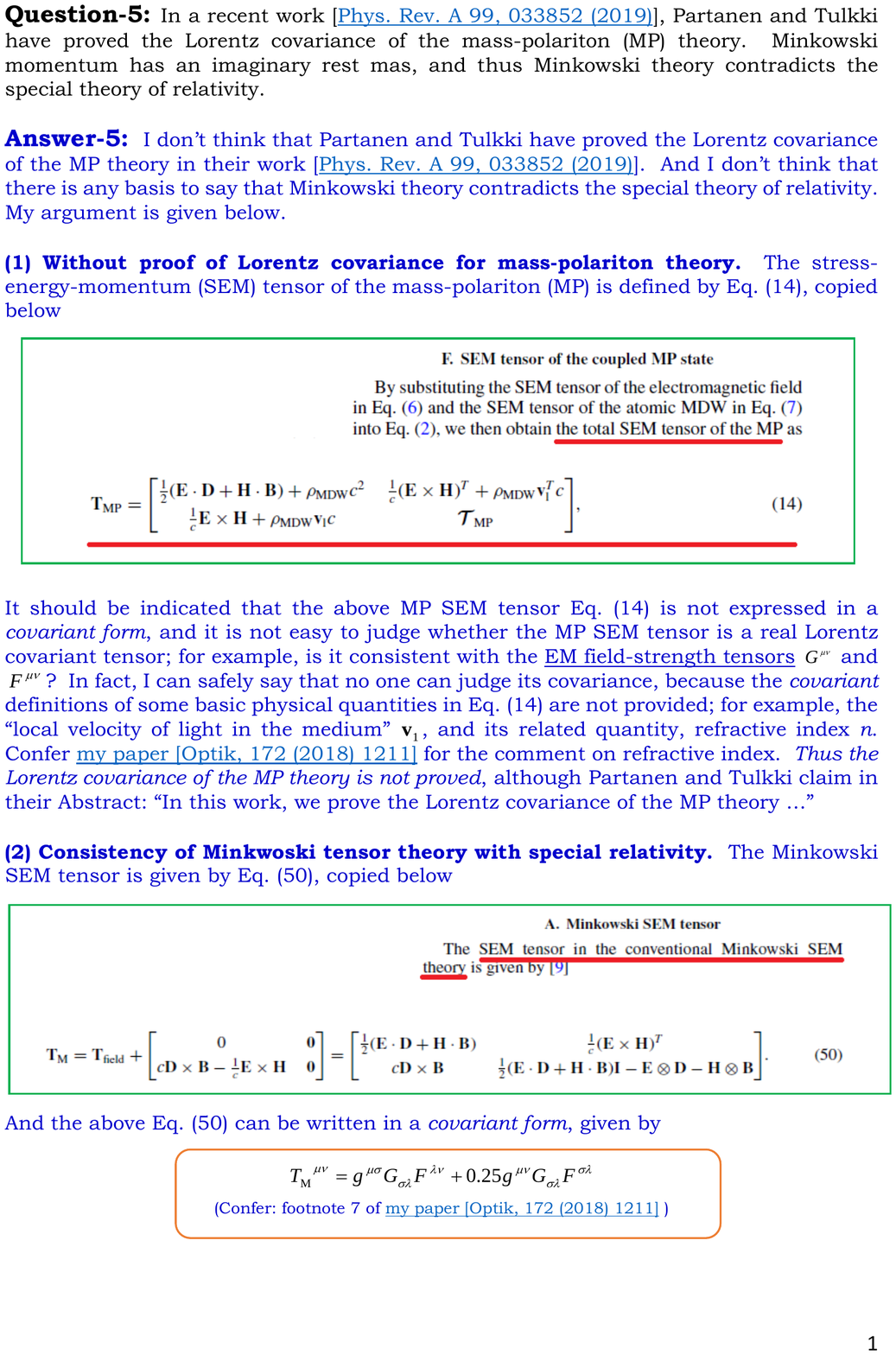}
\label{figM7}
\end{figure} 

\begin{figure} 
\includegraphics[trim=1.0in 1.0in 1.0in 1.0in, clip=true,scale=1.0]{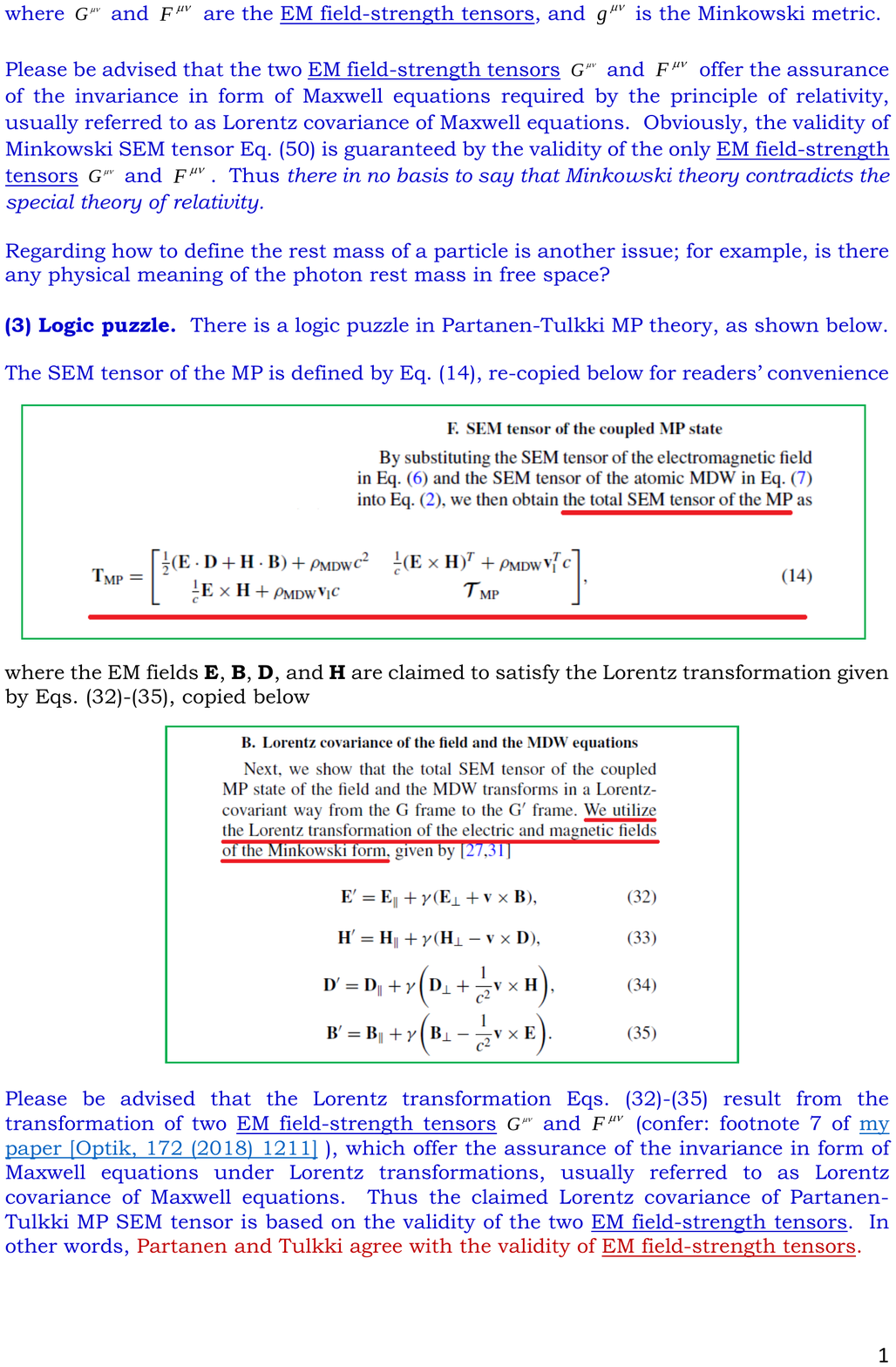}
\label{figM8}
\end{figure} 

\begin{figure} 
\includegraphics[trim=1.0in 1.0in 1.0in 1.0in, clip=true,scale=1.0]{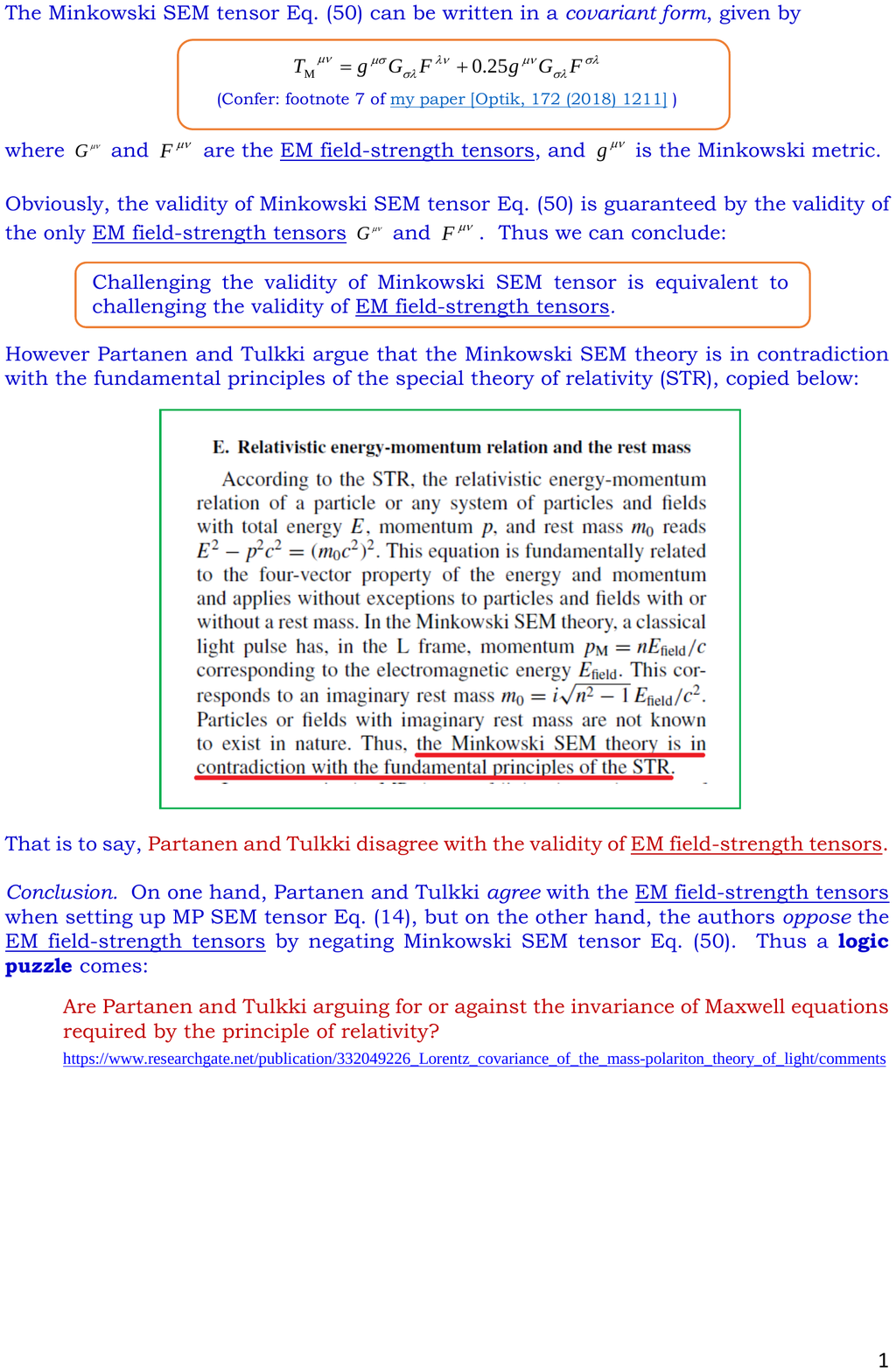}
\label{figM9}
\end{figure} 

\begin{figure} 
\includegraphics[trim=1.0in 1.0in 1.0in 1.0in, clip=true,scale=1.0]{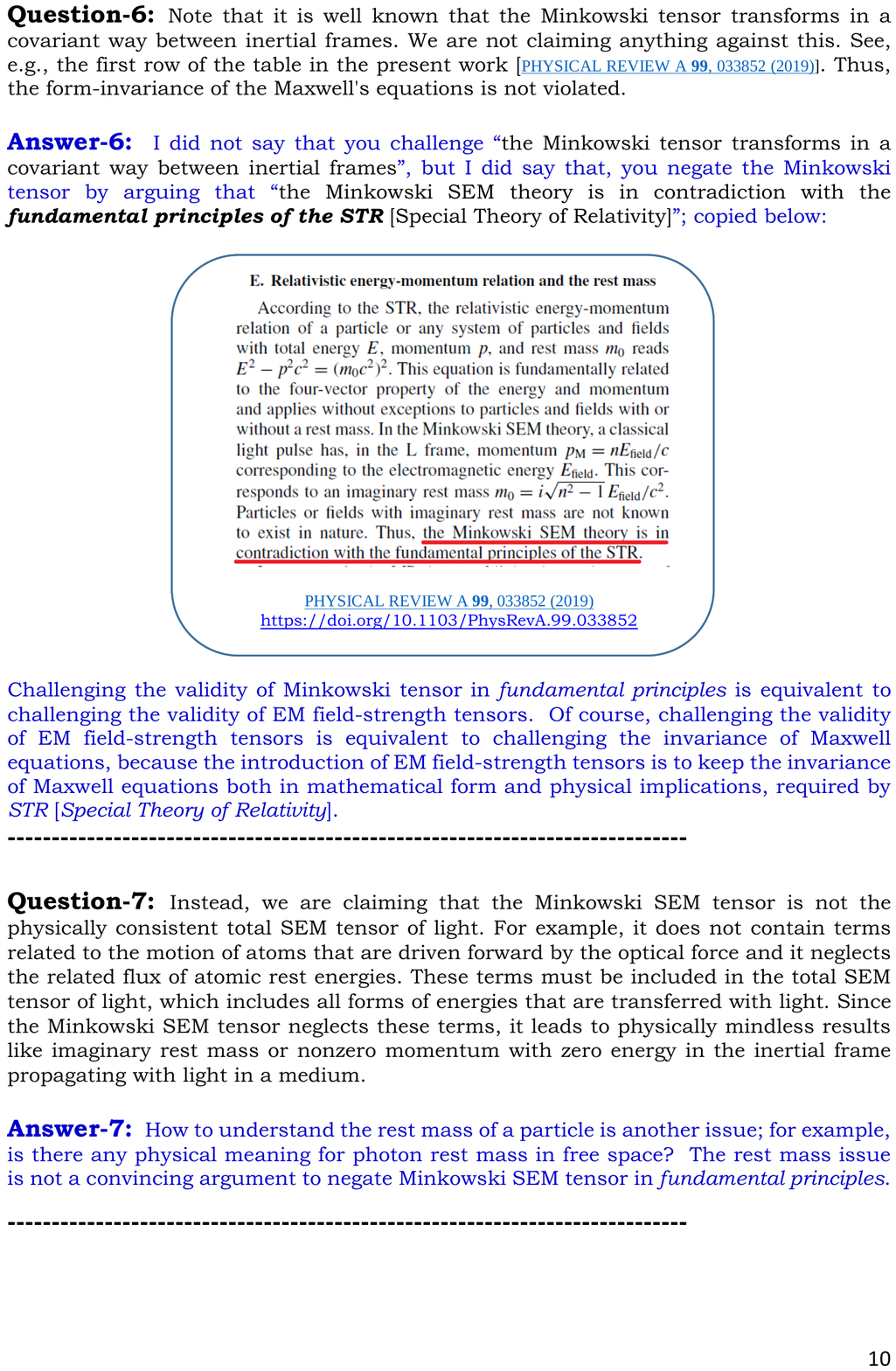}
\label{figM10}
\end{figure} 

\begin{figure} 
\includegraphics[trim=1.0in 1.0in 1.0in 1.0in, clip=true,scale=1.0]{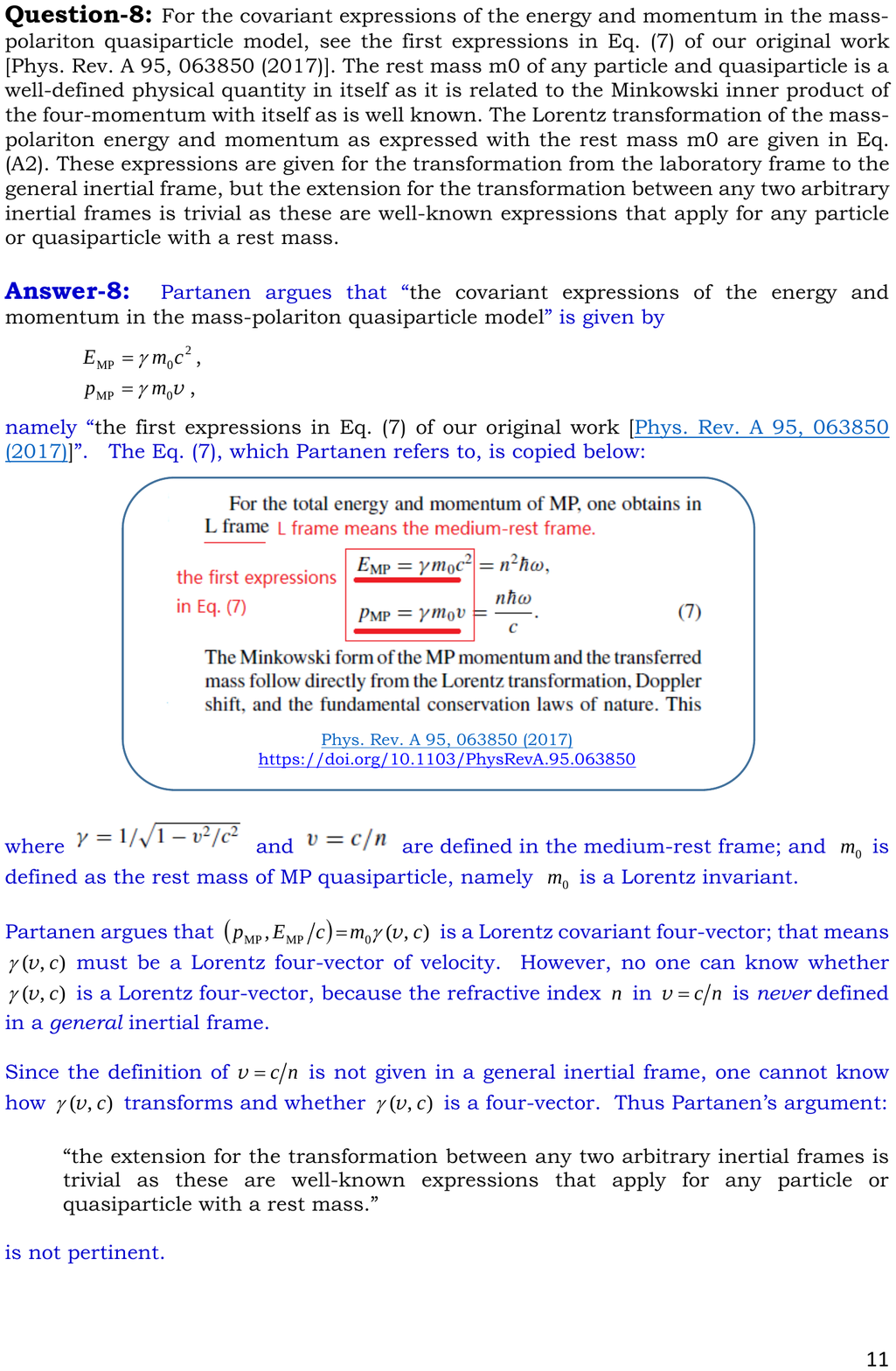}
\label{figM11}
\end{figure} 

\begin{figure} 
\includegraphics[trim=1.0in 1.0in 1.0in 1.0in, clip=true,scale=1.0]{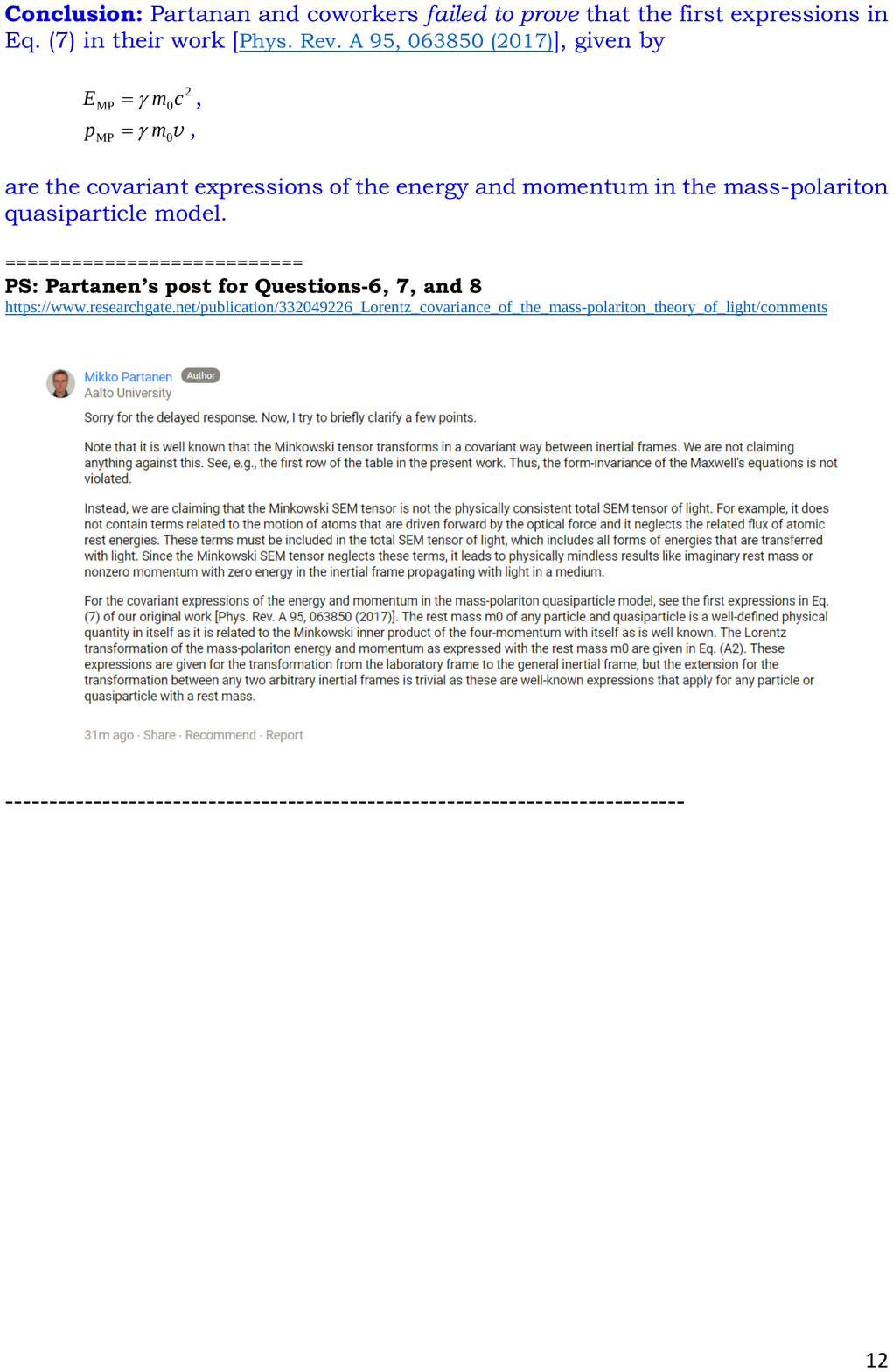}
\label{figM12}
\end{figure} 

\begin{figure} 
\includegraphics[trim=1.0in 1.0in 1.0in 1.0in, clip=true,scale=1.0]{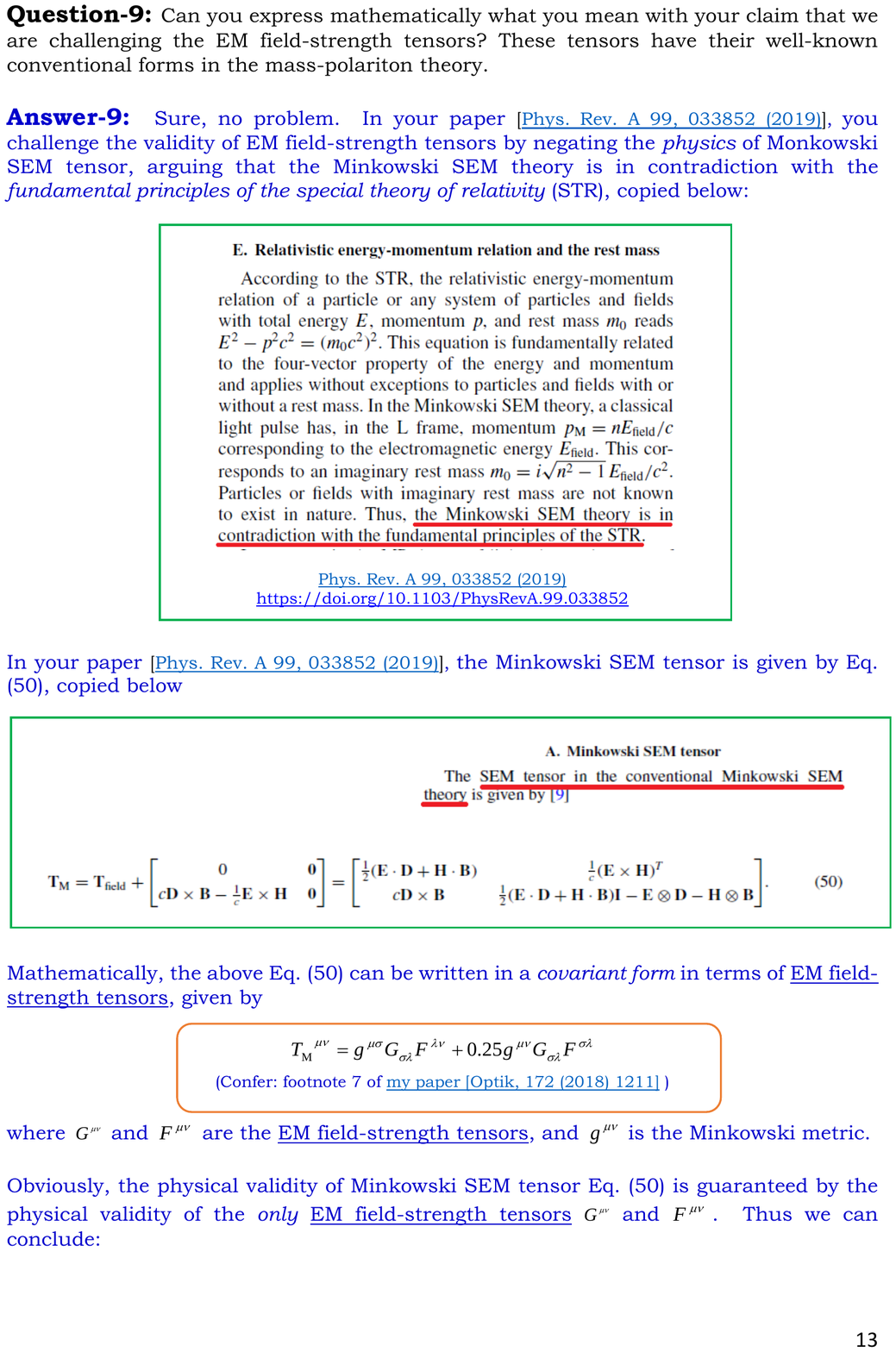}
\label{figM13}
\end{figure} 

\begin{figure} 
\includegraphics[trim=1.0in 1.0in 1.0in 1.0in, clip=true,scale=1.0]{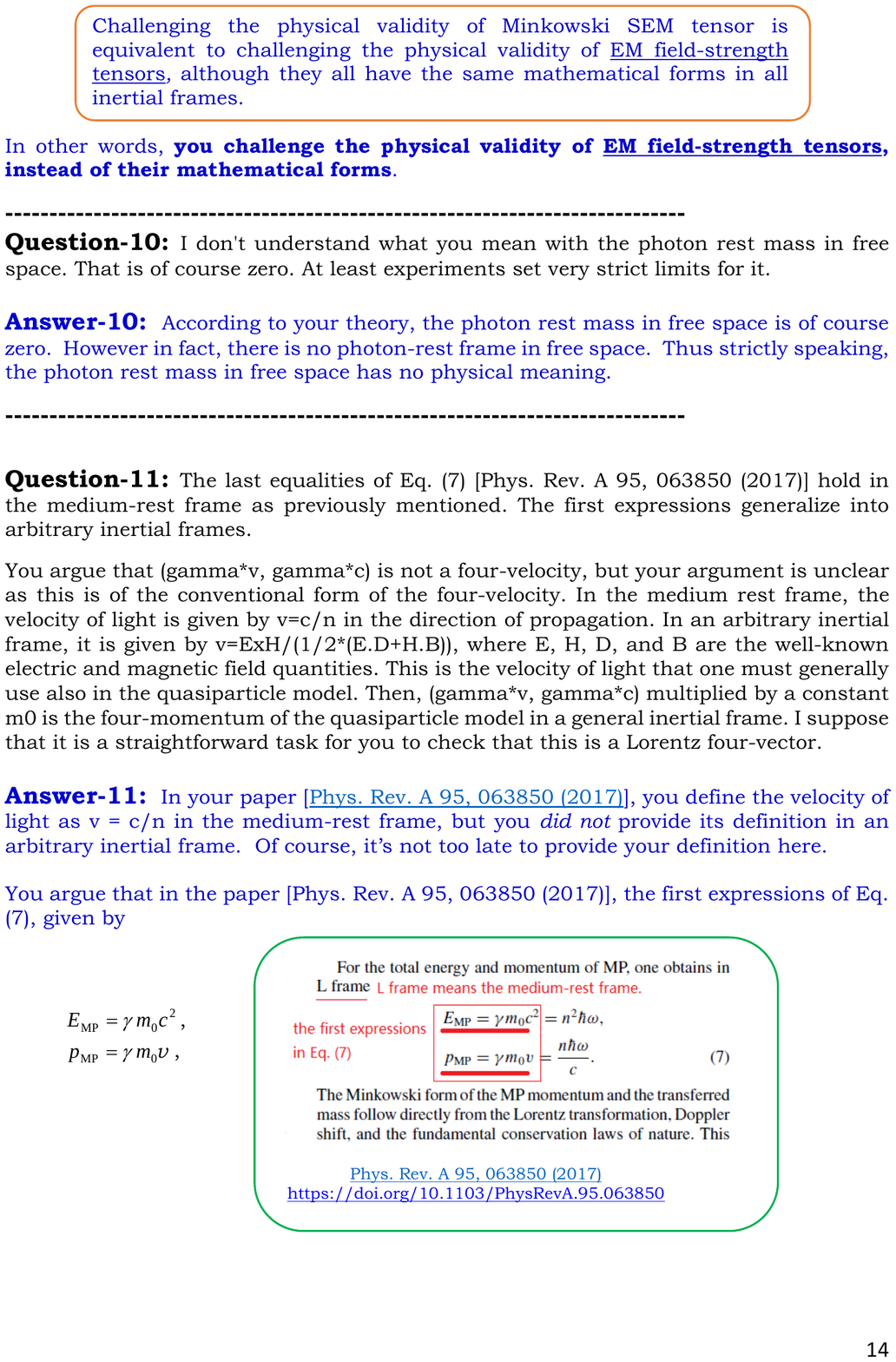}
\label{figM14}
\end{figure} 

\begin{figure} 
\includegraphics[trim=1.0in 1.0in 1.0in 1.0in, clip=true,scale=1.0]{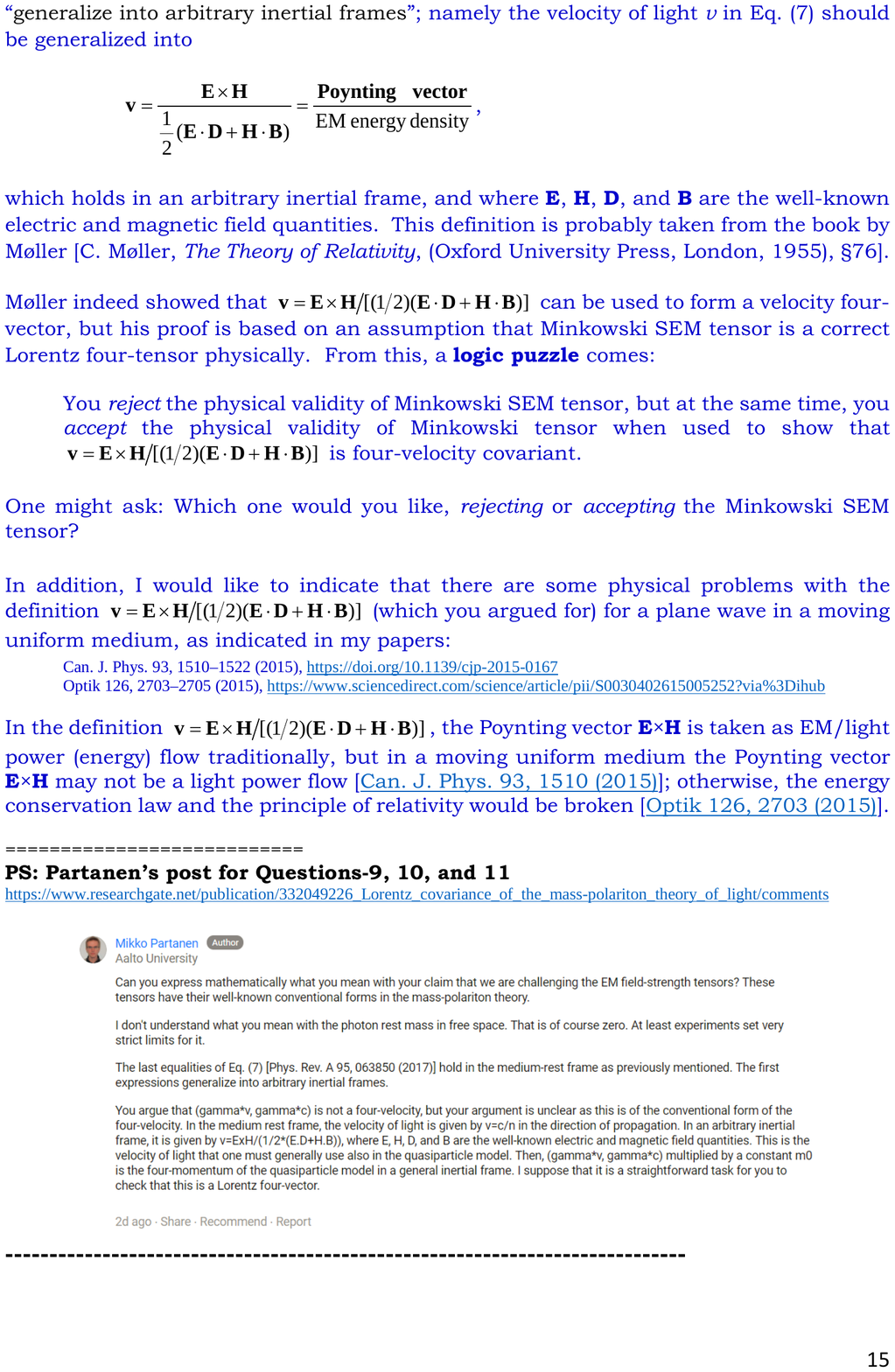}
\label{figM15}
\end{figure} 

\end{document}